\documentclass[pra,twocolumn,showpacs,amsmath,amssymb,superscriptaddress]{revtex4-1}

\usepackage{graphicx}
\usepackage{amsmath}
\usepackage{amssymb}
\usepackage{array}
\usepackage{SIunits}
\usepackage{color}
\usepackage{dcolumn}

\def\ca40{$^{40}\mathrm{Ca}^+$}
\def\T2{$\mathrm{T_2$}}
\def\Pr3{$\mathrm{Pr^{3+}}$}
\def\Eu3{$\mathrm{Eu^{3+}}$}
\newcommand{\EuYSO}[0]{Eu$^{3+}$:Y$_2$SiO$_5\,$}
\def\ket#1{$\left|#1\right>$}
\def\mket#1{\left|#1\right>}
\def\mbra#1{\left<#1\right|}
\def\mrm#1{$\mathrm{#1}$}

\def\fref#1{\ref{Fig:#1}}

\begin{document}

\title{High fidelity readout scheme for rare-earth solid state quantum computing}

\author{A.~Walther}\email[]{andreas.walther@fysik.lth.se} 
\author{L.~Rippe}
\author{Y.~Yan}
\author{J.~Karlsson}
\author{D.~Serrano}
\author{A.~N.~Nilsson}
\author{S.~Bengtsson}
\author{S.~Kr\"oll}
\affiliation{Department of Physics, Lund Univeristy, 221 00 Lund, Sweden}

\date{\today}

\begin{abstract}
We propose and analyze a high fidelity readout scheme for a single instance approach to quantum computing in rare-earth-ion-doped crystals. The scheme is based on using different species of qubit and readout ions, and it is shown that by allowing the closest qubit ion to act as a readout buffer, the readout error can be reduced by more than an order of magnitude. The scheme is shown to be robust against certain experimental variations, such as varying detection efficiencies, and we use the scheme to predict the expected quantum fidelity of a CNOT gate in these solid state systems. In addition, we discuss the potential scalability of the protocol to larger qubit systems. The results are based on parameters which we believed are experimentally feasible with current technology, and which can be simultaneously realized.
\end{abstract}


\maketitle

\section{Introduction}
The possibility of realizing a quantum computer is being investigated using a large variety of different experimental implementations. Currently, the largest entangled qubit systems have been realized in ion traps~\cite{Haeffner2005, Monz2010} and using linear optics with single photons~\cite{Yao2012}. There is however an intrinsic value in investigating solid state systems, as they are generally regarded as having a higher potential for future scalability to larger systems. For solid state systems, the best progress has been achieved with superconducting circuits~\cite{Devoret2013} and impurity-doped solids, such as NV centers in diamond~\cite{Wrachtrup2006}. Another impurity-doped system, rare-earth ions in crystals, have demonstrated very good performance in terms of quantum memories~\cite{Riedmatten2008,Afzelius2010,Hedges2010}, but has yet to demonstrate reliable two-qubit gates between spin qubits or a realistic route towards larger qubit systems. A major obstacle has been that, so far, only large ensembles of rare-earth ions have been used for gate operations~\cite{Fraval2005,Rippe2008}, and this has been shown to scale poorly~\cite{Wesenberg2007}. A promising approach to scalability in rare-earth quantum computing (REQC) is to move into the single instance regime, although this requires detecting single rare-earth ions inside their crystal hosts. Bare detection of single ions was just recently realized~\cite{Kolesov2012, Utikal2014a} with certainty, but there has been no clear description of how these dection schemes can be directly used in quantum information processing.

In this paper, we present a readout scheme that in principle allows for an arbitrarily high readout fidelity of the quantum state of a single ion inside a macroscopic host. The readout scheme is based on using a special buffer step (indicated in Fig.~\fref{ion_chain}) that can by cycled repeatedly, a scheme that is similar in its nature to what has been done previously for multi-species atomic clocks~\cite{Hume2007}. We show that with such a readout, a full Controlled-NOT (CNOT) gate can be performed in these systems with fidelity of about 99~\%, based on simulations that are supported by what is currently experimentally achievable. We also discuss further scaling towards larger multi-qubit systems by showing how chains of single ions can be mapped out, and we find that, including most known error sources as discussed in Sec.~\ref{cnot}, entangled states of 10 qubits remaining above 92~\% fidelity appear feasible, as long as all ions can control each other.

\begin{figure}[ht]
	\includegraphics[width=7.5cm,clip=true]{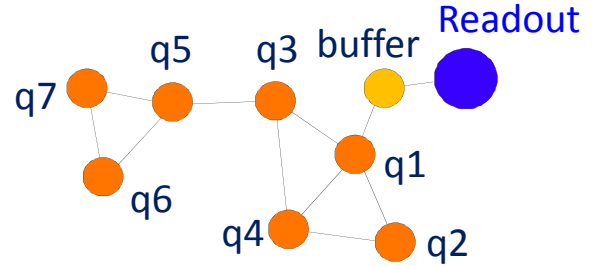} 
	\caption{(color online) A chain consisting of one readout ion surrounded by several qubit ions (e.g. Eu), where the closest qubit ion is being used as a buffer stage (the bright one), see the text for more details. The lines between the ions show which of them that can interact directly via frequency shifts caused by changes to the permanent dipole moments. With a 4~\% doping concentration, it is expected that each Eu ion can on average interact with about 5 other Eu ions surrounding it.}
	\label{Fig:ion_chain}
\end{figure}

It is interesting to note that at the single ion level, these impurity doped systems resembles the trapped ion systems but with two major difference. The first is the advantage that the ions are trapped by the comparatively large trapping potentials of the crystal bindings. This enables the ions to sit much closer to each other than in ion traps (nanometers instead of micrometers), which in turns allows the direct electrical dipole interactions between ions to be used as entangling mechanism. The second difference, is the disadvantage that the surrounding environment is not vacuum, but a crystal host that can cause additional decoherence effects as well as worsen the single ion detection possibilities through e.g. scattering. While the disadvantages may at first appear daunting, it is important to note that one of the main limitations to scalability in ion traps is that the entangling mechanism, the common motional modes, becomes increasingly more complex the more ions that are involved~\cite{Kielpinski2002}. The direct dipole interactions that can be used in REQC however, does not suffer from this problem, and we thus expect that once the initial hurdle of establishing single ion readout is overcome, the scaling to larger number of qubits will be much more manageable.

The paper is organized as follows: In Sec.~\ref{overview} the basic single instance quantum computing scheme is described together with a discussion on reasonable material parameters. Sec.~\ref{readout} details the readout scheme, and in Sec.~\ref{cnot} we go through a full CNOT gate from initialization to readout. Further scalability to larger qubit systems is discussed in Sec.~\ref{scalability} followed by a summary of our findings in Sec.~\ref{conclusions}.

\section{Overview and parameter considerations}
\label{overview}

The single instance scheme is based on using single rare-earth qubit ions that have suitable ground state hyperfine levels with long coherence times and in addition a long lived excited state that can be used for ion-ion interactions. Europium has generally demonstrated impressive coherence properties~\cite{Zhong2015}, and throughout the paper we will assume Eu as a qubit ion. It is very difficult to detect single ions with long lifetimes however, and to circumvent this, several schemes could be considered such as those where a readout ion of a different species is used. Coupling between ions, both between two qubit ions for gates and between a qubit ion and a readout ion for detection, will be mediated via permanent dipole-dipole interactions. In both cases, when two ions are sufficiently close to each other, the change of the static dipole moment as one ion is excited, is enough to shift the energy level of the neighboring ion out of resonance with a driving laser, thus providing a control mechanism.

Previously, a considerable attention has been given to using the short lived 5d transition in Ce as a potential readout ion of a different species~\cite{Walther2009b,Yan2013,Kolesov2013}. However, recent measurements have revealed that Eu absorbs at the same wavelength as the cycling transition in Ce, which makes it necessary to find an alternative readout ion. A very promising scheme for detecting single rare-earth ions is via Purcell enhancement of fluorescence due to coupling of the ion to a high finesse cavity with very small mode volume. A fiber based cavity setup~\cite{Kaupp2013} is a suitable candidate and would allow single ion detection of in principle any rare-earth 4f transitions. As an example we will here use Nd, which has a relatively high oscillator strength, but in case of unexpected energy transfers or overlapping absorption lines, any other rare-earth ion could be used with our readout scheme with no significant losses.

It will be assumed that we are working with a \EuYSO crystal, where 4~\% of the Yttrium ions in the crystal host have been replaced with Europium, distributed roughly equal in each of two different sites (though with only one isotope). This is a relatively high doping concentration and simulations have shown that, given the difference between the dipole moment of the ground and excited state, any ion will on average have more than 5 other ions sufficiently close to be controlled by it. For the readout ion on the other hand, background trace elements of Nd is expected to be enough, no special doping is required since one readout ion is enough for an entire chain of qubits. 

Any state-to-state transfer will be done with complex hyperbolic secant (\emph{sechyp}) pulses. These chirped pulses have the advantage over simple square pulses that they are robust against certain errors, such as amplitude and frequency fluctuations, see Ref.~\cite{Roos2004} for more details. Bloch simulations suggest that the Eu ions can be transferred to and from the excited state by such pulses of 400 ns duration with an efficiency of 99.96~\%, which will be used for the following calculations. The transfer efficiency is limited almost entirely by the duration of the pulse relative to the excited state lifetime, where the lower limit of the duration is set by the qubit hyperfine level separations. It should be noted that the transfer efficiency for Eu has not been fully verified by experiments, and does not include effects such as spectral diffusion. It is believed however, that the effects from spectral diffusion can be strongly mitigated by holeburning sequences that aims at keeping the total number of ions in the qubit frequency channels very low. The high transfer efficiency can be compared with experiments performed with the similar element Praseodymium, where the experimental transfer efficiency matches simulations rather well. For Praseodymium, the measured and calculated efficiency is about 96~\%~\cite{Rippe2008}, and the main limitations are the short excited state lifetime and the limited Rabi frequency available, as well as the fact that an ensemble was used that have not only an inhomogeneous frequency spread but also sits on different spatial parts of the beam profile, making different ions experience different Rabi frequencies. For single ion Eu transfers inside a cavity, both of those two limitations are strongly reduced, and preliminary measurements on Eu ensembles also supports that higher fidelities higher can be obtained in the Eu systems.

\section{Readout scheme}
\label{readout}

\begin{figure}[ht]
	\includegraphics[width=8.5cm,clip=true]{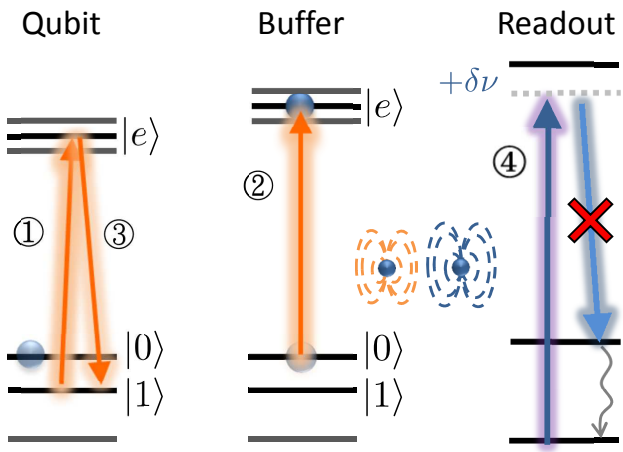} 
	\caption{(color online) Pulse sequence for reading out the state of one qubit via one buffer ion and the readout ion. In the instance shown the qubit is in \ket{0} state, and the first pulse, (1), resonant with the $\mket{1} \rightarrow \mket{e}$ qubit transition, does not excite it. The buffer ion is now unshifted, and a pulse, (2), resonant with the $\mket{0} \rightarrow \mket{e}$ buffer transition, will cause an excitation of the buffer ion. At this stage, the qubit ion is coherently returned to the ground state by pulse (3), such that it spends a minimum amount of time in the excited state. Finally, the readout ion is continuously excited and light is detected, (4). In the current example, the reaout ion was shifted due to an excited buffer ion and there is no fluorescence, but if the qubit was originally in the other state, the buffer would not get excited and thus the readout ion would instead fluoresce.}
	\label{Fig:pulse_seq}
\end{figure}

The state of the qubit ions can be read out with a readout ion using a permanent dipole blockade mechanism, which is also used for the quantum gates~\cite{Ohlsson2002,Wesenberg2007}. This mechanism is demonstrated in Fig.~\fref{pulse_seq}. In order to determine whether one qubit ion is in state \ket{0} or \ket{1}, it should be selectively exited to state \ket{e} with a pulse resonant with the $\mket{0} \rightarrow \mket{e}$ transition. If the qubit ion is excited, the readout ion's transition frequency is Stark shifted by the DC electric dipole field of the excited state of the qubit ion. This means that a readout laser tuned to the readout ion's unshifted resonant frequency, will not excite it.

The readout ion, in our example Nd, has a lifetime of 100 \micro s. With a reasonable cavity finesse of $10^4-10^5$ and a mode volume of a few wavelength's cubed~\cite{Kaupp2013}, a Purcell factor higher than $10^4$ can be achieved. Taking into account the decay branching ratios, we then obtain an effective readout ion lifetime of about 200 ns, which can thus be cycled many times during the duration of the qubit excitation, as \mrm{T_{1,Eu}} = 1.9 ms. The collection efficiency of a typical fluorescence detection setup may be about 1~\%, however, in a cavity with a high Purcell factor, almost all light will be spontaneously emitted into the same spatial mode. This can yield collections efficiencies in excess of 90~\%, and including other losses like detector quantum efficiency, can allow a total detection efficiency to go up to 10~\%. To demonstrate the flexibility of our readout scheme, scenarios with different collection efficiencies has been simulated using a Monte Carlo method, for both qubit starting states, \ket{0} and \ket{1}, as shown in Fig.~\fref{readout}. The blue histograms shows the number of collected photons when the readout ion is unshifted (qubit in state \ket{1} before the excitation pulse), and the red ones when it is shifted (qubit in state \ket{0} before the excitation pulse). The best results is achieved for different detection times depending on the total collection efficiency and background light level. For instance, for 1~\% collection efficiency, a collection time of 0.15\mrm{T_{1,Eu}} was found to be optimal with the probability to determine the correct state reaching approximately 93~\%. The largest source of error here is spontaneous decay of the qubit ion due to the finite lifetime of Eu. For a 10~\% collection efficiency, the optimal collection time was instead 0.025\mrm{T_{1,Eu}}.

\begin{figure}[ht]
	\includegraphics[width=8.5cm,clip=true]{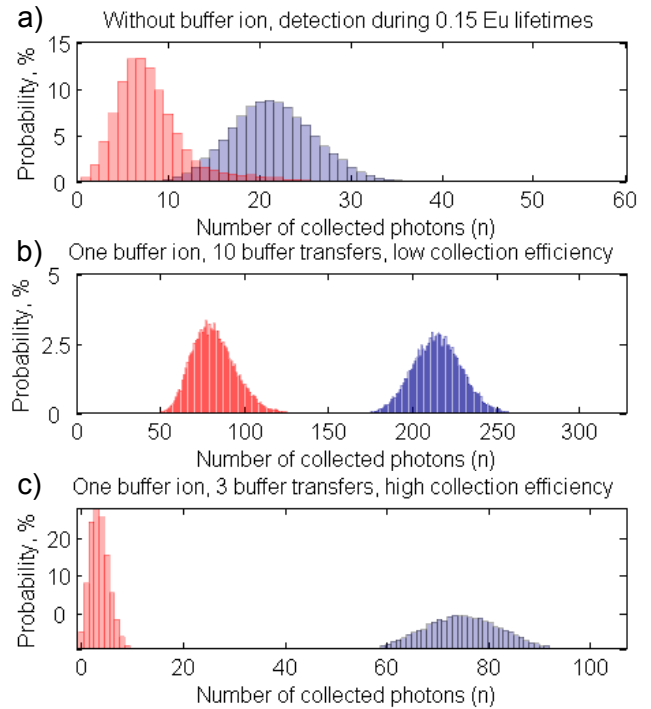} 
	\caption{(color online) Shows histograms of simulated photon statistics for detection of a single Nd ion in a cavity. In the upper panel Nd is directly controlled by the qubit ion. In the two lower panels the cumulative statistics from 10 and 3 repeated readouts respectively using one buffer step. The y-axis shows the probability of receiving $n$ photons during the optimal detection time, and the corresponding probabilities of correctly distinguishing the states is a) 93~\%, b) 99.7~\% and c) 99.85~\%. For the two lower panels, a detection efficiency of 1 and 10~\% respectively has been used, which allows fewer buffer transfers in the last panel.}
	\label{Fig:readout}
\end{figure}

The readout protocol can be improved significantly by introducing a buffer ion between the qubit ion and the readout ion, as shown in Fig.~\fref{pulse_seq}. For the discussion we use the same ion species for buffer as for qubit, but another species could potentially be used with some advantages. In this version, two mutually interacting qubit ions are used, where one of them, the buffer ion, can control the readout ion. Excitation of different qubit ions and the buffer ion can be distinguished by them having different resonance frequencies, i.e. they sit on different parts of the inhomogeneous profile. First, one pulse is used to state selectively excite the qubit ion, then another to selectively excite the buffer ion, conditioned on the qubit ion not being excited. A third pulse then coherently deexcites the qubit ion back to the ground state, if it was excited during the earlier stage (inverse of the first pulse). After this selective excitation the buffer ion state is read out as described above, and the number of photons from the readout ion during the 0.15\mrm{T_{1,Eu}} detection time is counted. This pulse sequence is illustrated in Fig.~\fref{pulse_seq}.

After one detection event, the buffer ion can be reinitilized through optical pumping. The qubit ion is still in its original state, and the same pulse sequence can be applied again to make another fluorescence measurement, yielding further information about the same qubit ion state. This process can be repeated several times, such that the total effective number of detected photons that depends on the qubit state can be increased substantially. The optimal number of times to repeat this buffering sequence depends on the detection efficiency and the background light level of the particular setup. Panels b) and c) of Fig.~\fref{readout} shows two different setups with 1~\% and 10~\% detection efficiency respectively that both can reach a probability of readout out the correct qubit state of about 99.7-99.9~\%. We note that roughly the same state distinguishability can be reached for both cases, showing that the scheme is robust against such experimental parameters, but that a larger number of buffer transfers is needed for lower detection efficiency. The final error is given mostly by the amount of time the qubit ion spends in the excited state, which cannot be reduced lower than the time it takes to do a state transfer on the buffer ion (see Sec.~\ref{cnot} for more details). In principle further buffer stages could be concatenated for an exponentially decreasing error probability, however, the buffer state transfer time relative to the Eu excited state lifetime makes protocols with more than one buffer stage unrewarding for Eu in particular (but could still be usefull for other setups).

\section{CNOT gate fidelity}
\label{cnot}

A full CNOT gate experiment will include the following steps (also confer Fig.~\fref{pulse_seq}):

\begin{enumerate}
	\item Initialization
	\item $\pi/2$-pulse, between $\mket{0}$ and $\mket{1}$, on the control qubit
	\item $\pi$-pulse, $\mket{0} \rightarrow \mket{e}$, on control qubit
	\item NOT on target qubit
	\item $\pi$-pulse reversing the excitation in step 3
	\item Readout
\end{enumerate}

The different steps in the list above will now be described in detail, including assumptions and expected errors for each step. The total CNOT error obtained in the end will include the error from all steps, with care taken to model the different nature of the errors. For example, any transfer pulse will cause both bit and phase flip errors, while any time spent in the excited state will be subject to lifetime decay, modeled as an amplitude damping channel (see e.g. Nielsen and Chuang~\cite{Nielsen2000a}).

\emph{Initialization - } The initialization step starts with finding a suitable chain of ions that can function as buffer and qubits, and can be described in 4 main steps: (i) find a fluorescing readout ion. (ii) scan through the inhomogeneous width of the qubit ions until the fluorescence from the readout stops, at which point an ion sufficiently close to the readout ion to shift it in frequency has been found. This will be the buffer ion. (iii) The inhomogeneous width of the qubit ions is scanned from start again in a pulsed manner. For each frequency channel a $\pi$-pulse is applied followed by a pulse exciting the buffer ion, and monitoring when the readout ion resumes fluorescing. This indicates that an ion that can control the buffer ion, preventing it from being excited, has been found. (iv) repeat the previous step one more time, such that two qubit ions that are both in the vicinity of the buffer ion is found. They will most likely also be sufficiently close to each other, but if they are not, and longer chains of ions is desired, the step is instead extended to find ions that shift the previous qubit ion, thus stopping it from controlling the buffer ion. This process can be nested as many layers away from the readout ion as it takes, with the overhead cost of only one extra pulse per layer away. 

After a sufficient chain of controlling ions has been established, the qubit ions should be initialized to the \ket{0} state, which can be done by means of optical pumping to an auxiliary state followed by a state transfer back in a similar manner as protocols used previously in the ensemble approach~\cite{Rippe2008}. The error in this step is assumed to be equal to the error of the final transfer pulse (a sechyp, see Sec.~\ref{overview}), i.e. the starting state is considered to be a mixed state with a probability of being in the wrong level of $4 \cdot 10^{-4}$ for each qubit.

\emph{$\pi/2$-pulse, between $\mket{0}$ and $\mket{1}$, on the control qubit - } Transitions between the hyperfine states cannot be directly driven by an optical laser field. Instead, arbitrary single qubit gates can e.g. be performed by two bichromatic pulses using a dark state technique (as described in Ref.~\cite{Rippe2008}). Such pulses have the same duration as the transfer pulses (defined in Sec.~\ref{overview}), and thus essentially have the same error. Since two pulses are needed, this step will give phase and bit flip errors twice as large as that of a transfer pulse, i.e. $8 \cdot 10^{-4}$ on the control qubit.

\emph{$\pi$-pulse, $\mket{0} \rightarrow \mket{e}$, on control qubit - } This step is a straightforward sechyp pulse, with phase and bit flip errors both considered to be $4 \cdot 10^{-4}$ for the control qubit.

\emph{NOT on target qubit - } Although the target qubit operation is conditioned on the control ion not causing a frequency shift, this step is essentially just a $\pi$-pulse on the hyperfine levels, i.e. a single qubit gate, making an error on the target qubit equal to $8 \cdot 10^{-4}$. In addition, the control qubit spends two pulse durations of time in the excited state, which gives a decay probability due to limited lifetime of $1-e^{-1.6 \mathrm{\mu s}/1.9 \mathrm{ms}} \approx 8 \cdot 10^{-4}$.

\emph{$\pi$-pulse reversing the excitation in step 3 - } Same operation and errors as the step 3 excitation.

\emph{Readout - } For the purpose of finding the achievable CNOT fidelity we use the scheme with one buffer ion, as described above, using a readout error of $2 \cdot 10^{-3}$ that we obtained earlier. Note that this error is asymmetrical, i.e. it represents the probability that a \ket{0} is counted as \ket{1}. The reverse error is usually much smaller which is because the main error is decay from the excited state, and state \ket{1} is never excited.

\emph{Final experiment fidelity - } The effects of all operations described above is calculated from actions applied to a starting density matrix. In the end when all steps have been taken into account, but before the readout, the system will be in a final density matrix, $\rho_f$. We can then compute the fidelity of the state as $F=\mbra{\psi_{max}}\rho_f\mket{\psi_{max}}$, where $\psi_{max}$ is the state we aim to create, such as a maximally entangled Bell state. Without the readout step, the total error $\varepsilon = 1 - F$ is found to be $ \sim 7 \cdot 10^{-3}$. The readout is included by allowing the calculated density matrix to be sampled as it would during a real readout sequence, with projections to the four possible two-qubit states. A quantum state tomography sequence was then simulated, using 15 different observables in 9 different measurement settings (see e.g.~\cite{Roos}). This gives a recreated density matrix (shown in Fig.~\fref{cnot}) that can be used to obtain the fidelity, including the readout stage, and we find the total error to be $ \sim 1 \cdot 10^{-2}$. This means that both the coherent operations and the readout process contribute significantly to the overall fidelity, which emphasizes the need for using the proposed readout buffer stage.

\begin{figure}[ht]
	\includegraphics[width=8.5cm,clip=true]{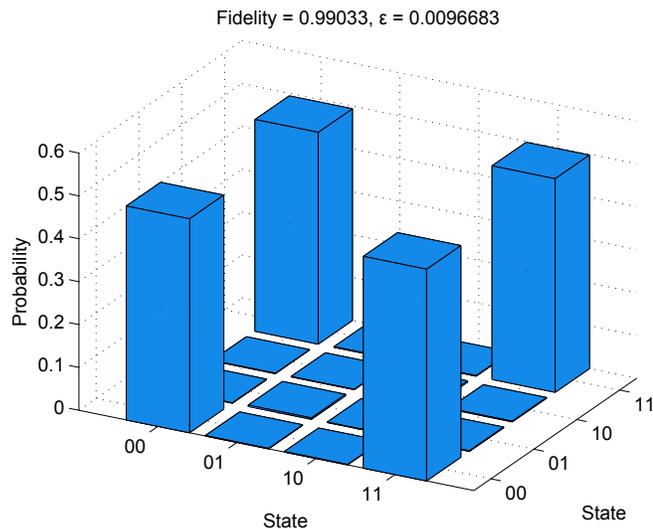} 
	\caption{(color online) Shows the (real) elements of the density matrix of a prepared Bell state, including all error sources as described in the text. The total fidelity without readout is 99.3~\% and with readout 99.0~\%.}
	\label{Fig:cnot}
\end{figure}

Note that while an effort has been made to include most systematic error sources, a real experiment will also include random projection noise caused by a limited number of experimental count cycles, but this has not been included here. In practice, this error will be limited by the total duration of the full protocol, and since gate operations can be made in sub-microseconds, the largest time is consumed by the readout stage. However, the fluorescence detection itself is not the main culprit, as a buffer stage with 10 repetitions, each one with a detection time of 0.15\mrm{T_{1,Eu}} or better will last a maximum of 3 ms, comparable to other single ion detection rates. Instead, the main time consumption arises from the reinitialization of the buffer step that has to be done between each repetition cycle. If simple optical pumping via the long lived excited state is used to reset the buffer state, then several lifetimes of Eu has to be used to reset it with good fidelity, which is a few tens of ms per repetition. To circumvent this, a quenching mechanism can be used by means of stimulating the transition from the excited state down to another Stark level, which then decay very fast to the ground state by non-radiative processes.

\begin{figure}[ht]
	\includegraphics[width=8.5cm,clip=true]{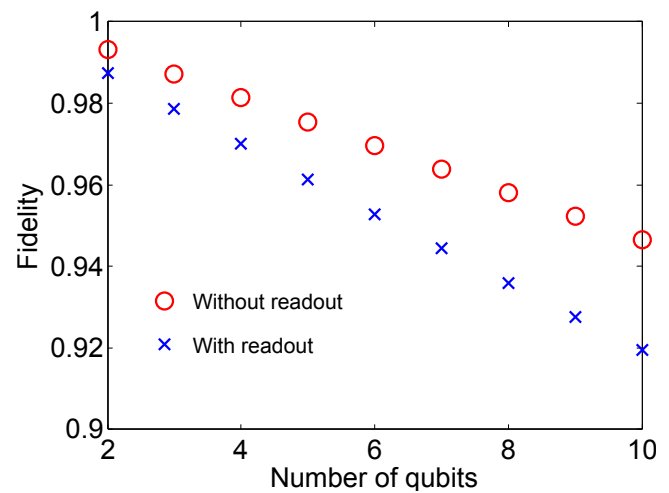} 
	\caption{(color online) Predicted fidelity of n-qubit GHZ states with and without readout. The fidelity includes pulses for tomography but assumes that all ions can interact with each other, which is reasonable at least up to 5 qubits with 4~\% doping concentration.}
	\label{Fig:nGHZ}
\end{figure}

\section{Scalability}
\label{scalability}

The previous section detailed the specific case of a two-qubit gate, where the errors were included in a careful manner in the total density matrix describing the system. This approach is difficult to extend to larger qubit systems, as the size of the Hilbert space scales exponentially with the number of qubits. In this section, we will attempt to give some figures for the scaling of larger qubit states by simpler considerations, based on the values of the CNOT gate obtained in the previous section. We will focus on the expected fidelity of an $n$-qubit Greenberger-Horne-Zeilinger (GHZ) state of the form $\mket{\Psi} = \mket{0...0} + \mket{1...1}$, which is a simple yet useful type of entangled state. The expected fidelity of this state can be fairly straightforwardly calculated by realizing that it is created by $n-1$ successive CNOT gates. Moreover, this fidelity will be the same both for the case of only nearest-neighbor interactions and for the case where each ion can control each of the other ions. This result is shown in Fig.~\fref{nGHZ}, both with and without readout. One limitation of the prediction is that while the creation of the GHZ state allows situations where only nearest neighbor interactions are possible, the readout step is calculated with the assumption that each ion can control a buffer ion directly without additional swap operations. As discussed earlier in Sec.~\ref{overview}, this is expected to be a reasonable case for at least up to 5 qubits for a doping concentration of 4~\%. The obtained fidelities indicate that the single instance rare-earth quantum computer schemes can be comparable to those of other multi-qubit schemes, such as trapped ions~\cite{Monz2011} or superconducting qubits~\cite{Barends2014}.

\section{Conclusions}
\label{conclusions}

We have presented a readout scheme for detecting quantum states of single ions inside a crystal host. The scheme is based on monitoring the cavity-enhanced fluorescence from one rare-earth ion using another long lived rare-earth ion species as a buffer stage that can be repeatedly cycled. Several buffer stages can be concatenated to yield a very long effective detection times, such that readout errors can be reduced more than one order of magnitude and reach $\varepsilon = 10^{-3}$ for a wide variety of collection efficiencies and background levels. We then used this result together with known error sources to obtain expected fidelities for a CNOT gate of 99~\% and for larger GHZ states remaining above 92~\% for up to 10 qubits. One of the limitations of our assumptions is presently that the expected increase in performance for qubit rotations when switching from Pr to Eu has not been fully experimentally verified as of yet. Our results indicate that rare-earth quantum computing can be feasible in the single instance regime.

\section*{Acknowledgments}
The authors acknowledge useful discussions with David Hunger. This work was supported by the Swedish Research Council, the Knut \& Alice Wallenberg Foundation, the Crafoord Foundation. The research leading to these results also received funding from the People Programme (Marie Curie Actions) of the European Union's Seventh Framework
Programme FP7 (2007-2013) under REA grant agreement no. 287252 (CIPRIS), Lund Laser Center (LLC), and the Nanometer Structure Consortium at Lund University (nmC@LU).


\bibliographystyle{apsrev4-1}
\bibliography{cnot}

\end{document}